\documentclass{acm_proc_article-sp}
\usepackage{mathrsfs,color,url,cite,citesort}
\newtheorem{theorem}{Theorem}
\newtheorem{definition}{Definition}
\newtheorem{lemma}{Lemma}
\newtheorem{corollary}{Corollary}
\usepackage[ruled]{algorithm2e}
\usepackage[utf8]{inputenc}
\usepackage[T1]{fontenc}

\def\real{\mathbb{R}}

\begin{document}

\title{Sketched SVD: Recovering Spectral Features from Compressive Measurements\titlenote{Copyright statement goes here.}}

\numberofauthors{3}
\author{
\alignauthor
Anna C. Gilbert \\
       \affaddr{University of Michigan}\\
       \affaddr{Ann Arbor, Michigan}\\
       \email{annacg@umich.edu}
\alignauthor
Jae Young Park \\
       \affaddr{University of Michigan}\\
       \affaddr{Ann Arbor, Michigan}\\
       \email{jaeypark@umich.edu}
\alignauthor
Michael B. Wakin \\
       \affaddr{Colorado School of Mines}\\
       \affaddr{Golden, Colorado}\\
       \email{mwakin@mines.edu}
}

\maketitle
\begin{abstract}
We consider a streaming data model in which $n$ sensors observe individual streams of data, presented in a turnstile model.  Our goal is to analyze the singular value decomposition (SVD) of the matrix of data defined implicitly by the stream of updates.  Each column $i$ of the data matrix is given by the stream of updates seen at sensor $i$.  Our approach is to sketch each column of the matrix, forming a ``sketch matrix'' $Y$, and then to compute the SVD of the sketch matrix. We show that the singular values and right singular vectors of $Y$ are close to those of $X$, with small {\em relative} error. We also believe that this bound is of independent interest in non-streaming and non-distributed data collection settings.

Assuming that the data matrix $X$ is of size $N \times n$, then with $m$ linear measurements of each column of $X$, we obtain a smaller matrix $Y$ with dimensions $m \times n$. If $m = O(k \epsilon^{-2} (\log(1/\epsilon) + \log(1/\delta))$, where $k$ denotes the rank of $X$, then with probability at least $1-\delta$, the singular values $\sigma'_j$ of $Y$ satisfy the following relative error result
\[
(1-\epsilon)^{\frac{1}{2}}\leq\frac{\sigma'_j}{\sigma_j} \leq (1 + \epsilon)^{\frac{1}{2}}
\]
as compared to the singular values $\sigma_j$ of the original matrix $X$.  Furthermore, the right singular vectors $v'_j$ of $Y$ satisfy
\begin{align*}
&\|v_j-v_j'\|_2\leq\nonumber\\
&\min\left\{\sqrt{2}, ~ \frac{\epsilon\sqrt{1+\epsilon}}{\sqrt{1-\epsilon}}\max_{i\neq j}\frac{\sqrt{2}\sigma_i\sigma_j}{\displaystyle\min_{c\in[-1,1]}\{|\sigma^2_i-\sigma^2_j(1+c\epsilon) |\}  } \right\},
\end{align*}
as compared to the right singular vectors $v_j$ of $X$. We apply this result to obtain a streaming graph algorithm to approximate the eigenvalues and eigenvectors of the graph Laplacian in the case where the graph has low rank (many connected components).
\end{abstract}


\section{Introduction}\label{sec:intro}

Consider a collection of data arranged in a matrix $X$ of size $N \times n$. Each column represents a length-$N$ signal (or image, frequency counts of terms in a particular document, etc.) and there are $n$ such observations. The singular value decomposition (SVD) of $X$,
\[
	X = U \Sigma V^T,
\]
carries important information about the structure of the data set, especially when the rank $k$ of $X$ is small. In particular, the columns of $U$ (known as the {\em left} singular vectors of $X$) span the principal directions of the data set and can be used as basis vectors for building up typical signals, and the diagonal entries of $\Sigma$ (known as the singular values of $X$) reflect the energy of the data set in each of these directions. The extraction of these features is commonly known as Principal Component Analysis (PCA), and PCA is a fundamental and commonly used tool in data analysis and compression. This exact same process can be viewed through a slightly different lens when one imagines the columns of $X$ as independent realizations of a length-$N$ random vector $x$. Computing the left singular vectors of $X$ is equivalent to computing the eigenvectors of $X X^T$, which (up to rescaling) is the $N \times N$ sample covariance matrix of the data. In this context, PCA is also known as the Karhunen-Loeve (KL) Transform, and the KL Transform is a fundamental and commonly used tool in statistics.

There are in fact a number of applications where the {\em right} singular vectors $V$ of a data matrix $X$ are more important, or equivalently, where the eigenvectors of $X^T X$ carry the structure of interest. For example, the product $\Sigma V^T$ gives a low-dimensional embedding of the data set that preserves distances and angles between the $n$ data vectors. This embedding can be used for clustering or categorizing the signals \cite{ng2002spectral,belkin2001laplacian}; for example, this is used for comparing documents in latent semantic analysis. The right singular vectors of $X$ can also be viewed as the result of applying the KL Transform to the rows, rather than the columns, of $X$. In this sense, the columns of $V$ describe the inter-signal (rather than intra-signal) statistical correlations. In cases where the column index corresponds to a distinct sensor position, or a vertex in a graph, etc., this correlation structure can carry important structural information \cite{spielman2009spectral,chung1997spectral}.

Unfortunately, many of the applications in which we seek the right singular vectors of $X$ (equivalently, the eigenvectors of $X^TX$) are those in which the data is simply too large, too distributed, or generated too quickly for us to store the data or to process it efficiently in one, centralized location. There are, however, settings in which the data sets---while large---have low intrinsic dimension or are of low rank.  Let us suppose that the length of each data vector $N$ is much larger than the number of observations $n$, and suppose that $X$ has rank $k \leq n$.  The data may or may not be generated in a dynamic, streaming fashion and it may or may not be collected in a distributed fashion amongst $n$ sensors.

We wish to design a joint observation process (which can be distributed amongst $n$ sensors) that maintains a ``sketch'' of the data stream and a reconstruction process that, at a central location, reconstructs {\em not} an approximation of the original data, but rather a good approximation to the singular values $\sigma_j$ and the (right) singular vectors $v_j$ of the original matrix $X$.  The sketch of the data stream should be a linear, non-adaptive procedure, one that is efficient to update, and one that uses as few observations of the data matrix as possible so that as little communication as possible is required from the sensors to the central processing entity.  Because the procedure is linear and non-adaptive, we can represent the sketch as a matrix-matrix product $\Phi X = Y$, where the observation matrix $\Phi$ is of size $m \times N$ and the sketch $Y$ is of size $m \times n$.  We want $m$ as small as possible.  From the sketch matrix $Y$, we want to produce estimates $\sigma_j'$ of the $k$ non-zero singular values and estimates $v_j'$ of the associated (right) singular vectors of $X$ such that
$$
(1-\epsilon)^{1/2} \sigma_j \le \sigma_j' \leq (1 + \epsilon)^{1/2} \sigma_j
$$
and
$$
	\| v_j-v_j'\|_2 \leq \epsilon \gamma,
$$
where $\gamma$ is a function of the smallest gap between the singular values of $X$.

Most of the current work on low-rank matrix approximations, robust PCA~\cite{candes2011robust}, rank-revealing QR factorizations (see \cite{HalkoMartinssonTropp:2011} and the references therein for a comprehensive survey), etc.\ has focused on obtaining a good approximation $\widehat{X}$ to the original data matrix $X$, albeit one that is parsimonious.  Some of the measurement schemes~\cite{candes2011robust} sketch both the row and column space of $X$, collecting measurements $Y = \Phi X \Phi^T$, and some sketch the column space only so as to derive a few orthogonal vectors that span the column
space~\cite{HalkoMartinssonTropp:2011}.  These results are of the form
\[
	\| X - \widehat{X}\|_{\mathcal X} \leq (1 + \epsilon) \|X\|_{\mathcal X}
\]
for some norm ${\mathcal X}$ (typically, the Frobenius norm, but others as well).  Our goal is an approximation to the {\em singular vectors} and {\em singular values} of $X$ themselves, directly, and while one could apply standard perturbation theory techniques to compare the singular vectors of an approximation $\widehat{X}$ to those of the original data $X$, the error guarantees would be rather poor. Furthermore, while one could ask about preserving the subspaces spanned by the singular vectors, there are many applications from PCA to data clustering, image segmentation, graph embedding, and modal analysis in which the individual singular vectors are critical for data analysis tasks, and data reconstruction is not necessarily required.

Our algorithmic approach is straightforward. We sketch one side of the $N\times n$ data matrix $X$, maintaining a sketch matrix $Y = \Phi X$ of size $m \times n$. (The fact that the sketch is one-sided allows it to be computed sensor-by-sensor in distributed data collection settings.) We then compute the SVD of the sketch matrix $Y$, using standard (iterative) SVD algorithms.
Our analysis is quite different from that of most randomized linear algebra methods. We assume that the sketching matrix $\Phi$ is randomly generated and satisfies the distributional Johnson-Lindenstrauss (JL) property (see Definition~\ref{def:jl}) so that with high probability it acts as a near isometry on the column span of $X$, and we then exploit {\em relative error} (as opposed to absolute error) perturbation analysis for {\em deterministic} (as opposed to random) matrices to obtain our results.
As detailed in our main result, Theorem~\ref{thm:mainResult02}, we can obtain accurate relative estimates for the singular values, and in some cases we can obtain accurate estimates for the singular vectors as well. However, we struggle to achieve high accuracy in the singular vectors when the corresponding singular values are close. 
This is a consequence of well-studied perturbation theory and seems inherent in our approach.

One major application of our work is to determining structural graph properties of streaming graph data, albeit for low-rank graphs (ones with many connected components). Recent work on the structure and evolution of online social networks~\cite{KumarNovakTomkins:2010} suggests that a significant fraction of vertices in such networks participate in isolated communities, ``small groups who interact with one another but not with the network at large.''

In Section 2, we set the stage for our mathematical problem and in Section 3, we outline the related work, including an overview of the relative error perturbation techniques from linear algebra that we will use.  In Section 4, we present our main result which we apply, in Section 5, to the spectral analysis of streaming graphs.

\section{Problem Setup}\label{sec:problemsetup}

Let $X$ denote the $N\times n$ real-valued data matrix. For example, one may envision that each column represents a length-$N$ time series signal collected from one of $n$ sensors.
Let us assume that $N\geq n$, and that $\text{rank}(X)=k\leq n$.
We write the \textit{truncated} SVD of $X$ as $X = U_X \Sigma_X V_X^T$, where unlike the full SVD, $U_X$ is $N\times k$, $\Sigma_X=\text{diag}(\sigma_1,\dots,\sigma_k)$ with $\sigma_1 \ge \cdots \ge \sigma_k > 0$, and $V_X$ is $n\times k$.
Our goal in this paper is to estimate the singular values $\Sigma_X$ and the right singular vectors $V_X$ from a low-dimensional sketch obtained by left-multiplying $X$ with a compressive matrix.
In particular, we let $\Phi$ denote a sketching matrix of dimension $m\times N$, and we denote the sketch of $X$ by $Y = \Phi X$.
We are specifically interested in cases where $m < N$ and thus $Y$ is a shorter matrix than $X$.
We also note that the sketching matrix $\Phi$ can be applied individually to each column of $X$, and these sketched columns can be concatenated to form the $m \times n$ matrix $Y$.
In other words, the sketching can be performed sensor-by-sensor.

Our algorithm for estimating $\Sigma_X$ and $V_X$ from $Y$ is explained in Section~\ref{sec:algorithm}.
This algorithm is very simple and is based on the idea that under a suitable choice of $\Phi$, the singular values and right singular vectors of $Y$ will approximate the singular values and right singular vectors of $X$.
In order to state our results, we write the truncated SVD of $Y$ as $Y = U_Y \Sigma_Y V_Y^T$, where $U_Y$ is $m\times k$, $\Sigma_Y=\text{diag}(\sigma'_1,\dots,\sigma'_k)$ with $\sigma'_1 \ge \cdots \ge \sigma'_k \ge 0$, and $V_Y$ is $n\times k$.
(We will be interested in cases where $m \ge k$, and typically $Y$ will have rank $k$ just like $X$.)
It will also be useful for us to write the eigendecompositions of $X^T X$ and $Y^T Y$ as $X^T X=V_X\Lambda_X V_X^T$ and $Y^T Y=V_Y\Lambda_Y V_Y^T$, respectively, where $\Lambda_X=\text{diag}(\lambda_1,\dots,\lambda_k)$ and $\Lambda_Y=\text{diag}(\lambda_1',\dots,\lambda_k')$.
(Although it is not common practice we will use the ``truncated'' eigendecomposition so that $V_X$ and $V_Y$ are both of dimension $n\times k$ and $\Lambda_X$ and $\Lambda_Y$ are both of dimension $k\times k$; when $k=n$ we will have the usual eigendecomposition.)
Note that $\lambda_j=\sigma_j^2$ and $\lambda_j'=\sigma'^{2}_j$ for $j=1,\dots,k$.

To ensure that the spectral information about $X$ is preserved in the sketched matrix $Y = \Phi X$, we rely on randomness to construct the sketching matrix $\Phi$.
Any random distribution that can be used to construct a Johnson-Lindenstrauss (JL) embedding can be used to generate $\Phi$.
\begin{definition}
\label{def:jl}
An $m \times N$ random matrix $\Phi$ is said to satisfy the {\em distributional JL property} if for any fixed $x \in \real^N$, and any $0 < \epsilon < 1$,
$$
\text{Pr}\left[ \left| \| \Phi x \|_2^2 - \| x \|_2^2 \right| > \epsilon \| x \|_2^2 \right] \le 2 e^{-m f(\epsilon)},
$$
where $f(\epsilon) > 0$ is a constant depending only on $\epsilon$.
\end{definition}

For most random matrices satisfying the distributional JL property, the functional dependence on $\epsilon$, $f(\epsilon)$, is quadratic in $\epsilon$ as $\epsilon \rightarrow 0$.
For compactness, we suppress this except where necessary for quantifying the number of measurements or the run time of our algorithm. There are a variety of random matrix constructions known to possess the distributional JL property. Notably, random matrices populated with independent and identically distributed (i.i.d.)\ subgaussian entries will possess this property~\cite{Daven_Concentration}. Subgaussian distributions include suitably scaled Gaussian and $\pm 1$ Bernoulli random variables.
Other examples of non-i.i.d.\ JL matrices and discussions of the function $f(\epsilon)$ are contained in works such as~\cite{Jayram:2011:OBJ:2133036.2133037,Dasgupta:2010:SJL:1806689.1806737}.
There are many papers that address the sparsity of a JL matrix, the speed of its application to a vector, the minimum number of rows it must possess, the minimum amount of randomness necessary to generate such a matrix, etc.  For our results, a random matrix satisfying the distributional JL property is sufficient and, depending on the particular application (streaming versus static data), we want either fast update times (i.e., a sparse transform) or a fast transform. We appeal to a long line of work in assessing the qualities of such transforms and in constructing them, either randomly or deterministically.

Finally, we note that a primary objective of this work is to quantify the amount of perturbation of the right singular vectors of $X$ under the random measurement operator $\Phi$. For $j=1,\dots,k$, let us denote the $j$th column of $V_X$ as $v_j$ and the $j$th column of $V_Y$ as $v'_j$.
Our bounds concern the quantity $\|v_j-v_j'\|_2$.
However, we note that the right singular vectors $v_j$ and $v_j'$ are each unique only up to multiplication by $-1$.
Thus, without loss of generality, we will assume that the sign of each $v_j'$ is chosen so that it is positively correlated with $v_j$. That is, we will assume for each $j$ that $\langle v_j, v_j'\rangle \geq 0$.

\section{Related Work}\label{sec:relatedwork}

In order to quantify the amount of change in the singular values and the right singular vectors, we approach this problem from the matrix perturbation theoretic perspective.
To see the connection to matrix perturbation theory, let us write
\begin{equation*}
Y^T Y = X^T\Phi^T\Phi X = V_X \Sigma_X U_X^T \Phi^T\Phi U_X \Sigma_X V_X^T.
\end{equation*}
Defining $\Delta_{\Phi}:=\Phi^T\Phi-I$, $Y^TY$ becomes
\begin{align*}
Y^TY&=V_X \Sigma_X U_X^T (I+\Delta_{\Phi}) U_X \Sigma_X V_X^T,\\
&=V_X \Sigma_X^2 V_X^T +V_X \Sigma_XU_X^T\Delta_{\Phi} U_X \Sigma_X V_X^T.
\end{align*}
Given this equation, we can think of the symmetric matrix $Y^TY$ as being the summation of some original symmetric matrix
$$
A:=V_X \Sigma_X^2 V_X^T
$$
and a perturbation matrix
$$
E:=V_X \Sigma_XU_X^T\Delta_{\Phi} U_X \Sigma_X V_X^T.
$$
Roughly speaking, when $E$ is small in some sense, it would be reasonable to expect $Y^TY$ to have approximately the same spectral information as $A$.
Thus we can think about our problem as quantifying the amount of change between the eigenvalues and eigenvectors of $A$ (which equal the squared singular values and the right singular vectors of $X$) and those of $Y^TY$ under the additive perturbation $E$.

In the following sections, we briefly review some of the important results in the matrix perturbation theory literature and also discuss the connection of our problem to the simultaneous iteration method, an important algorithm for computing the eigendecomposition of a matrix.

\subsection{Absolute Bounds}
There is an extensive literature in the field of matrix perturbation theory quantifying the amount of change in the eigenvalues and eigenvectors of a symmetric $n\times n$ matrix $A$ when it undergoes an additive perturbation.
A perturbed matrix, $\widetilde{A}$, may be written in the form of $\widetilde{A}=A+E$, where $E$ is the perturbation matrix that is being added to $A$.
It is well-known that the eigenvalues of $A$ and those of $\widetilde{A}$ will be close to one another when the amount of perturbation $E$ is small (typically with respect to the 2-norm of $E$).
Let us denote the $j$th largest eigenvalues of $A$ and $\widetilde{A}$ as $\lambda_j$ and $\widetilde{\lambda}_j$, respectively.
Then, it is known~\cite{golub1996} that for $j = 1,\dots,n$,
\begin{equation}
|\widetilde{\lambda}_j-\lambda_j|\leq \|E\|_2.
\label{eqn:abseigval}
\end{equation}
Thus we can see that the distance between each perturbed eigenvalue and the corresponding original eigenvalue will depend on the amount of perturbation, i.e., $\|E\|_2$.

To discuss the perturbation in the eigenvectors let us first represent the eigenvectors of $A$ as $v_j$ such that $Av_j=\lambda_j v_j$.
Similarly, let us define the eigenvectors of $\widetilde{A}$ as $\widetilde{v}_j$ such that $\widetilde{A}\widetilde{v}_j=\widetilde{\lambda}_j\widetilde{v}_j$.
It is well known that for general matrices $A$ and $E$, the eigenvectors $v_j$ and $\widetilde{v}_j$ may vary drastically even when the amount of perturbation is small.
In other words, $\|\widetilde{v}_j-v_j\|_2$ can be large even when $\|E\|_2$ is small.
To see why, let us for example look at the case when two eigenvalues, $\lambda_1$ and $\lambda_2$, of $A$ are equal to each other.
For such a case, we know that the eigenvectors corresponding to those eigenvalues, $v_1$ and $v_2$, will not be unique: any linear combination of the two eigenvectors will also be a valid eigenvector corresponding to the same eigenvalue.
A perturbation to this matrix will generally cause $\lambda_1$ and $\lambda_2$ to split into two eigenvalues $\widetilde{\lambda}_1$ and $\widetilde{\lambda}_2$, each of which will satisfy equation~\eqref{eqn:abseigval} above.
If $\widetilde{\lambda}_1$ and $\widetilde{\lambda}_2$ are distinct, the corresponding eigenvectors $\widetilde{v}_1$ and $\widetilde{v_2}$ will now be uniquely identified.
Since $v_1$ and $v_2$ were not unique, it is possible for $\widetilde{v}_1$ to differ from any particular choice of $v_1$ and for $\widetilde{v}_2$ to differ from any particular choice of $v_2$.

The perturbation in the eigenvectors, however, is not completely arbitrary.
It is known that the angle between the space spanned by $v_1$ and $v_2$ and the space spanned by $\widetilde{v}_1$ and $\widetilde{v}_2$ will be small if $E$ is small.
To state this result more concretely, let us represent the eigendecompositions of $A$ and $\widetilde{A}$ as
\begin{equation*}
A=V\Lambda V^{T}=
\left( V_1~V_2 \right)
\left(
\begin{array}{cc}
\Lambda_1 & 0\\
0 & \Lambda_2
\end{array}
\right)
\left(
\begin{array}{c}
V_1^{T} \\
V_2^{T}
\end{array}
\right)
\end{equation*}
and
\begin{equation*}
\widetilde{A}=\widetilde{V}\widetilde{\Lambda} \widetilde{V}^{T}=
\left( \widetilde{V}_1~\widetilde{V}_2 \right)
\left(
\begin{array}{cc}
\widetilde{\Lambda}_1 & 0\\
0 & \widetilde{\Lambda}_2
\end{array}
\right)
\left(
\begin{array}{c}
\widetilde{V}_1^{T} \\
\widetilde{V}_2^{T}
\end{array}
\right).
\end{equation*}
The eigenvalue matrices are such that,
\begin{align*}
&\Lambda_1=\text{diag}(\lambda_1,\dots,\lambda_p),~\Lambda_2=\text{diag}(\lambda_{p+1},\dots,\lambda_n),\\
&\widetilde{\Lambda}_1=\text{diag}(\widetilde{\lambda}_1,\dots,\widetilde{\lambda}_p),~\widetilde{\Lambda}_2=\text{diag}(\widetilde{\lambda}_{p+1},\dots,\widetilde{\lambda}_n)
\end{align*}
for an arbitrarily chosen $p \in \{2,\dots,n-1\}$.

It is possible to quantify how close the spaces spanned by the columns of $V_1$ and $\widetilde{V}_1$ are.
In order to provide a measure of closeness between the two spaces, the following notion of angle matrix was defined in~\cite{davis1970}:
\begin{align*}
&\Theta(X_1,X_2):=\nonumber\\
&\arccos\left( (X_1^{T}X_1)^{-\frac{1}{2}} X_1^{T} X_2(X_2^{T}X_2)^{-1} X_2^{T}X_1 (X_1^{T}X_1)^{-\frac{1}{2}}   \right)^{-\frac{1}{2}},
\end{align*}
where $X_1$ and $X_2$ are two matrices of the same dimension $n\times p$ with $n>p$ and full column rank.
The singular values of $\Theta(X_1,X_2)$ are the angles required to rotate the space spanned by $X_1$ onto that of $X_2$.
Going back to our notation of $V_1$, $V_2$, $\widetilde{V}_1$, and $\widetilde{V}_2$, it was shown~\cite{davis1970} that
\begin{equation*}
\|\sin\Theta(V_1,\widetilde{V}_1)\|=\|\widetilde{V}_2^{T}V_1\|
\end{equation*}
for any unitarily invariant norm.
This fact can be used to bound the angle between $V_1$ and $\widetilde{V}_1$.
In particular, if $\kappa:=\min|\lambda(\Lambda_1)-\lambda(\widetilde{\Lambda}_2)|>0$, then
\begin{equation*}
\|\sin\Theta(V_1,\widetilde{V}_1)\|_F=\|\widetilde{V}_2^{T}V_1\|_F\leq \frac{\|E\|_F}{\kappa}.
\end{equation*}
Once again we note that the above bound relies on the absolute separation between eigenvalues (in contrast with the relative separation, which appears in Section~\ref{sec:relbounds}).
The above can be further generalized to any invariant norm.
Detailed discussion on this subject can be found in~\cite{davis1970}.

\subsection{Relative Bounds}
\label{sec:relbounds}

The perturbation results discussed above are in terms of the absolute differences between eigenvalues.
These types of results are most useful for ensuring the preservation of the largest eigenvalues but least useful for ensuring the relative preservation of the smallest eigenvalues; a small {\em absolute} change in a small eigenvalue could actually correspond to a large {\em relative} change in that eigenvalue.
The absolute perturbation results are the best we can do when the perturbation to $A$ is completely arbitrary.
However, when the perturbation exhibits some structure one can do much better than what the absolute error bounds indicate.

Consider the class of perturbations that take the form $\widetilde{A}=A+E=D^T A D$, where $D$ is non-singular.
It was shown~\cite{eisenstat1995} that in this case a relative perturbation bound for the eigenvalues is given by
\begin{equation*}
|\widetilde{\lambda}_j-\lambda_j|\leq |\lambda_j|\|D^T D- I\|_2,
\end{equation*}
where the factor $\|D^T D- I\|_2$ represents how close $D$ is to being an orthonormal matrix.
In the extreme case when $D$ has orthonormal columns we will have that $\widetilde{\lambda}_j=\lambda_j$ as expected.

Similarly, the angle $0\leq\theta_j\leq\frac{\pi}{2}$ between the $j$th eigenvector and its corresponding perturbed eigenvector has been shown~\cite{eisenstat1995} to satisfy
\begin{equation*}
\sin\theta_j\leq\frac{\|D^TD\|_2\|(DD^T)^{-1} -I\|_2}{\rho_j(A)-\|D^TD-I\|_2}+\|D-I\|_2,
\end{equation*}
provided that $\rho_j(A)>\|D^TD-I\|_2$,
where the $j$th relative gap, $\rho_j(A)$, of the eigenvalues of $A$ is defined as
\begin{equation*}
\rho_j(A)=\min_{i\neq j}\frac{|\lambda_i-\lambda_j|}{|\lambda_j|}.
\end{equation*}
We can see that for the type of perturbation described above we obtain a much stronger perturbation result that depends on the relative gap between the eigenvalues.
There are many variants of relative perturbation results that have been proposed to date~\cite{li98,li96,eisenstat1995,Barlow80computingaccurate} that differ from one another depending on the underlying matrix $A$ (e.g., whether it is a symmetric matrix, positive definite matrix, indefinite matrix, etc.) and also on the type of perturbation.

\subsection{Relation to Simultaneous Iteration}

Our problem also has a close connection to various algorithms for computing the eigenvalues and eigenvectors of symmetric matrices.
One algorithm that we shall focus on is the simultaneous iteration method~\cite{Rutishauser1969,Clint1970}.
This method is best suited for cases when we are interested in computing the top few eigenvalues and their corresponding eigenvectors and when the underlying matrix is sparse.

To state the algorithm explicitly, let us set some notation.
Let $A$ be an $n\times n$ positive-semidefinite matrix with eigendecomposition $A=V\Lambda V^T$. We let $\lambda_j$ and $v_j$ denote the $j$th eigenvalue of $A$ and its corresponding eigenvector.
We also assume that the eigenvalues are ordered in descending order such that $\lambda_1\geq\lambda_2\geq\dots\geq\lambda_n\geq0$.
We then pick a set of trial vectors and denote them as $p_1,p_2,\dots,p_k$, where the number of trial vectors $k$ depends on how many eigenvectors we wish to compute.
The trial vectors can be any set of orthonormal vectors such that
\begin{equation}
\label{eqn:nullspacecondition}
\text{span}(p_1,\dots,p_k)\bigcap \text{span}(v_{k+1},\dots,v_{n})=\{0\}.
\end{equation}
One possible choice of trial vectors is a set of $k$ orthonormal vectors that are chosen randomly.
Let us stack the trial vectors into columns of a matrix and denote it as $P=[p_1,\dots,p_k]$.
Given this notation the simultaneous iteration method is carried out as follows:\footnote{There are a few variants of the simultaneous iteration method. Here, we use the algorithm presented in~\cite{Rutishauser1969}.}
\vspace{-0.1in}
\begin{enumerate}
\item $W^{(0)} \leftarrow AP$
\item for $i=1,2,\dots$
\begin{enumerate}
\item $Q^{(i)}R^{(i)} \leftarrow W^{(i)}$ via the QR-decomposition
\item $W^{(i+1)} \leftarrow AQ^{(i)}$
\item if stopping criterion is not met, set $i \leftarrow i+1$ and go back to step $(a)$, otherwise output $Q^{(i)}$.
\end{enumerate}
\end{enumerate}
\vspace{-0.1in}
As we can see, the simultaneous iteration method iteratively refines the set of eigenvectors of $A$.
Denoting the $j$th column of $Q^{(i)}$ as $q_j^{(i)}$, it is known~\cite{Rutishauser1969} that
\begin{equation*}
\|q_j^{(i)}-v_j\|_2=\mathcal{O}(p_j^{i}),
\end{equation*}
where $p_j=\max\{\lambda_{j+1}/\lambda_j,\lambda_j/\lambda_{j-1}\}$.
From this we can see that the rate of convergence of the eigenvectors depends on the ratio between the eigenvalues.
Put differently, an eigenvector that corresponds to an eigenvalue with a favorable eigenvalue ratio $p_j$ will converge faster to the true eigenvector.
This is also similar in spirit to the relative perturbation results that we discussed above in that $p_j$ provides relative measure of the closeness between the eigenvalues, and the accuracy between $v_j$ and $q_j^{(i)}$ depends on $p_j$.

Lastly, let us focus on the very first step in the simultaneous iteration algorithm, in which we multiply the original matrix $A$ with $k$ (potentially randomly chosen) vectors.
Interestingly, this is similar in spirit to our algorithm, except that our data matrix $X$ is not necessarily square or positive-semidefinite, and we do not require the rows of $\Phi$ to be orthonormalized. 
We would like to note that---when they are used in the simultaneous iteration method---random trial vectors are merely chosen as a way to satisfy the condition~\eqref{eqn:nullspacecondition}.
We believe, however, that randomness will also help to better preserve the true eigenvectors in the first iteration.

\subsection{Randomized Algorithms for Linear Algebra}

In a similar vein, there have been a large number of results on what we will refer to as randomized algorithms for linear algebra.  The monograph~\cite{Mahoney:2011:RAM:2185807.2185808} covers a number of these methods and references.

There are several lines of work that are closely related to our results.
The first involves the spectral analysis of random matrices and the application to algorithmic tasks such as information retrieval and spectral random graph analysis.
Two representative works are \cite{Azar:2001:SAD:380752.380859,Dasgupta:2004:SAR:1032645.1033216} which build random matrix models (or random perturbations of random matrices) for data and graphs and then use those models to find the approximate spectral structures in the data (or the SVDs).
Many of the perturbation results used in these papers fall into our category of absolute bounds.

A second line of work is that of robust PCA or low-rank matrix completion, of which~\cite{candes2011robust} is just one example (there are many other such papers).
In this problem, the data matrix $X$ is either sampled or random linear measurements of the form $\Phi X \Phi^T$ are obtained, from which a sparse, low-rank approximation $\widehat{X}$ to the original data matrix is produced.
The primary goal of this line of work is to approximate the original data with a parsimonious representation.
Our work, in contrast, aims to recover or to approximate the parsimonious representation itself.

A third line of work is the recovery of principal components of a data matrix $X$ from compressive projection measurements~\cite{hughes2012,fowler2009}.
Briefly speaking, in these works, a rectangular data matrix $X$ of size $p\times n$ is considered where $p<n$, i.e., $X$ has more columns than rows.
Each column of $X$ represents a data sample, and the objective in these works is to compute the \textit{left} singular vectors of $X$ from compressive projections of the columns of $X$.
However, different random projection matrices are applied to different columns of $X$.
The key differences between the work proposed in~\cite{hughes2012,fowler2009} and our work are that 1) we are interested in the right singular vectors of $X$, 2) the data matrix $X$ in our problem is assumed to have more rows than columns, 3) our measurement matrix is a JL matrix, and 4) we apply the same random matrix to every column of $X$.

Finally, we emphasize the distinction between subspace approximation and the approximation of the singular vectors themselves.
Feldman et al.\ \cite{Feldman:2010:CSH:1873601.1873654} give coreset and sketching algorithms for approximating subspaces spanned by portions of the data set.
This problem is also similar to the work of Halko et al.\ \cite{HalkoMartinssonTropp:2011}, in which one constructs a basis for an approximate subspace spanned by the columns of $X$ from a sketch $Y = \Phi X$ of the data.
Finally, we note that the work of Drineas et al.\ \cite{Drineas:StatisticalLeverage:2011} that approximates leverage scores of a matrix is similar in nature to ours but does not produce approximate singular vectors.

\section{Main Result}\label{sec:mainresults}

\subsection{Proposed Algorithm and Estimation Bounds}
\label{sec:algorithm}

Recall the problem setup discussed in Section~\ref{sec:problemsetup}. Our algorithm for estimating $\Sigma_X$ and $V_X$ from the sketched matrix $Y = \Phi X$ is stated in Algorithm~\ref{algorithm}. This algorithm is very simple: we simply return the truncated singular values and right singular vectors of $Y$.

\begin{algorithm}
\label{algorithm}
	\KwIn{Sketched matrix $Y = \Phi X$}
	\KwOut{$\widehat{\Sigma}_X$ and $\widehat{V}_X$ (estimates of the singular values and right singular vectors, respectively, of $X$)}
	$(U_Y,\Sigma_Y, V_Y) \leftarrow {\rm SVD}(Y)$ \\
    $\widehat{\Sigma}_X \leftarrow \Sigma_Y$ \\
    $\widehat{V}_X \leftarrow V_Y$
\caption{Pseudo-code for sketched SVD.}
\end{algorithm}

The computational complexity of our algorithm can be divided into two parts.
The first part concerns the complexity of computing the sketch $Y$ from $\Phi$ and $X$.
If both $\Phi$ and $X$ are available at a central processing node, $Y$ can be computed simply by multiplying $\Phi$ and $X$; as discussed in Section~\ref{sec:problemsetup}, there may be a fast algorithm for doing this, depending on the structure of $\Phi$.
As we have noted, however, it is also possible to compute the sketch column-by-column by applying $\Phi$ separately to each column of $X$; when data is collected in a distributed fashion, this may be the natural way to construct a sketch.
Let us denote the computational complexity computing $Y$ as $T_1(m,N,n)$.
In distributed scenarios, we will have $T_1(m,N,n)=nT'(m,N)$, where $T'(m,N)$ denotes the computational complexity of one matrix-vector multiplication with $\Phi$.

The second part concerns the complexity of computing the SVD of $Y$.
Using standard techniques, computing the SVD of an $m\times n$ matrix with $m\geq n$ requires $\mathcal{O}(mn^2)$ operations.
(When $k$ is very small compared to $n$, we may have $m < n$ and the SVD of $Y$ can be computed even more efficiently than this.)
Combining this fact with the bound on $m$ provided in our main result (see~\eqref{eqn:m}) and assuming $f(\epsilon)$ is quadratic in $\epsilon$, the computational complexity of this second part will be $\mathcal{O}(n^2 k \epsilon^{-2} (\log(1/\epsilon) + \log(1/\delta))$, where $\delta$ denotes the failure probability.
One can add this cost to $T_1(m,N,n)$ to determine the overall computational complexity, although it is important to stress that the computation of $Y$ may be streaming or distributed over many sensors, while the computation of the SVD of $Y$ may be performed all at once at a central computing node.

We are now ready to state our main result.

\begin{theorem}
\label{thm:mainResult02}
Let $X$ be an $N \times n$ matrix with $N\geq n$ and $\text{rank}(X)=k\leq n$, and let $X= U_X \Sigma_X V_X^T$ denote the truncated SVD of $X$ as explained in Section~\ref{sec:problemsetup}.
Let $\epsilon\in (0,1)$ denote a distortion factor and $\delta\in(0,1)$ denote a failure probability, and suppose $\Phi$ is an $m \times N$ random matrix that satisfies the distributional JL property with
\begin{equation}
\label{eqn:m}
m\geq \frac{k\log(42/\epsilon)+\log(2/\delta)}{f(\epsilon/\sqrt{2})}.
\end{equation}
Let $Y = \Phi X$ denote the sketched matrix, and let $\widehat{\Sigma}_X = \Sigma_Y$ and $\widehat{V}_X = V_Y$ denote the estimated singular vectors and right singular values of $X$ returned by Algorithm~\ref{algorithm}.
Then with probability exceeding $1-\delta$, $\text{rank}(Y)=k$ and both of the following statements hold:
\begin{enumerate}
\item (Preservation of singular values) For all $j = 1,\dots,k$,
\begin{equation*}
(1-\epsilon)^{1/2}\leq\frac{{\sigma}'_j}{\sigma_j}\leq (1+\epsilon)^{1/2},
\end{equation*}
where $\sigma_1 \ge \cdots \ge \sigma_k \ge 0$ denote the singular values of $X$ (the diagonal entries of $\Sigma_X$) and $\sigma'_1 \ge \cdots \ge \sigma'_k \ge 0$ denote the singular values of $Y$ (the diagonal entries of $\widehat{\Sigma}_X$).
\item (Preservation of right singular vectors) For all $j = 1,\dots,k$,
\begin{align*}
&\|v_j-v_j'\|_2\leq\nonumber\\
&\min\left\{\sqrt{2}, ~ \frac{\epsilon\sqrt{1+\epsilon}}{\sqrt{1-\epsilon}}\max_{i\neq j}\frac{\sqrt{2}\sigma_i\sigma_j}{\displaystyle\min_{c\in[-1,1]}\{|\sigma^2_i-\sigma^2_j(1+c\epsilon) |\}  } \right\},
\end{align*}
where $v_1, \dots, v_k$ denote the right singular vectors of $X$ (the columns of $V_X$) and $v_1', \dots, v_k'$ denote the right singular vectors of $Y$ (the columns of $\widehat{V}_X$).
\end{enumerate}
\end{theorem}

{\bf Proof}: See Section~\ref{sec:proofs}.

\begin{corollary}
When $\Phi$ is generated randomly from an i.i.d.\ subgaussian distribution (suitably scaled) or some other random distribution satisfying the distributional JL property with quadratic $f(\cdot)$, the bounds in Theorem~\ref{thm:mainResult02} will hold with $m = \mathcal{O}(k \epsilon^{-2} (\log(1/\epsilon) + \log(1/\delta))$.
\end{corollary}

This result states that from $Y$ we can obtain accurate relative estimates for the singular values of $X$, and in some cases we can obtain accurate estimates for the right singular vectors of $X$ as well. However, we struggle to achieve high accuracy in the singular vectors when the corresponding singular values are close. This is a consequence of well-studied perturbation theory (recall the role that the relative gap played in Section~\ref{sec:relbounds}) and seems inherent in our approach.

We note that, naturally, we could obtain similar results for preserving the left singular vectors of $X$ were we to sketch its rows, rather than its columns. We also note that when the exact rank $k$ of the data matrix is unknown (or if the rank of $X$ is not necessarily below $n$), substituting $n$ for $k$ in the measurement bound \eqref{eqn:m} yields a guarantee that applies to any $N \times n$ matrix $X$ with $N \ge n$.

\subsection{Proof of Theorem~\ref{thm:mainResult02}}
\label{sec:proofs}

In this section we prove Statements 1 and 2 within Theorem~\ref{thm:mainResult02}.
As noted in Section~\ref{sec:problemsetup}, it will be useful for us to write the truncated eigendecompositions of $X^T X$ and $Y^T Y$ as $X^T X=V_X\Lambda_X V_X^T$ and $Y^T Y=V_Y\Lambda_Y V_Y^T$, respectively, where $\Lambda_X=\text{diag}(\lambda_1,\dots,\lambda_k)$ and $\Lambda_Y=\text{diag}(\lambda_1',\dots,\lambda_k')$.
Recall that $\lambda_j=\sigma_j^2$ and $\lambda_j'=\sigma'^{2}_j$ for $j=1,\dots,k$.

\subsubsection{Proof of Statement 1}
\label{sec:svalsproof}

In order to prove Theorem~\ref{thm:mainResult02}, we require the following result, adapted from Lemma 5.1 in~\cite{baraniuk2008simple} and Theorem 4.3 in~\cite{davenport2010random}.
\begin{lemma}[\cite{baraniuk2008simple,davenport2010random}]
\label{lem:DPSS}
Let $\mathcal{X}$ denote a $k$-dimensional subspace of $\mathbb{R}^N$.
Let $\epsilon\in (0,1)$ denote a distortion factor and $\delta\in(0,1)$ denote a failure probability, and suppose $\Phi$ is an $m \times N$ random matrix that satisfies the distributional JL property with
\begin{equation*}
\label{eqn:measurements}
m\geq \frac{k\log(42/\epsilon)+\log(2/\delta)}{f(\epsilon/\sqrt{2})}.
\end{equation*}
Then with probability exceeding $1-\delta$,
\begin{equation*}
\label{eqn:com}
\sqrt{1-\epsilon}\|x\|_2\leq \|\Phi x\|_2\leq \sqrt{1+\epsilon}\|x\|_2,
\end{equation*}
for all $x\in\mathcal{X}$.
\end{lemma}

To see how this lemma can help us guarantee the preservation of the singular values of $X$ (or, equivalently, the eigenvalues of $X^T X$), we begin by noting that
\begin{equation*}
Y^TY=X^T\Phi^T\Phi X=V_X\Sigma_X U_X^T\Phi^T\Phi U_X\Sigma_X V_X^T.
\end{equation*}
We define a new $k\times k$ matrix
$$
M:=V_X^T Y^TY V_X=\Sigma_X U_X^T\Phi^T\Phi U_X\Sigma_X
$$
and represent its eigendecompostion as $M=V_M\Lambda_M V_M^T$.
Noting that $Y^T Y=V_Y\Sigma_Y^2 V_Y^T$, we have
$$
M=V_X^T Y^TYV_X=V_X^T V_Y \Sigma_Y^2 V_Y^TV_X.
$$
From this we can infer that $\Lambda_M=\Sigma_Y^2$, i.e., $M$ has the same set of eigenvalues as $Y^T Y$.
Thus, we turn our attention to proving that the eigenvalues of $M$ approximate the eigenvalues of $X^T X$.
Let us define $\Delta_{\Phi}:=\Phi^T\Phi - I$ and consider the ratio
\begin{align*}
\frac{x^T M x}{x^T \Sigma_X^2 x}&=\frac{x^T \Sigma_XU_X^T\Phi^T\Phi U_X\Sigma_Xx}{x^T \Sigma_X^2 x}\nonumber\\
&=\frac{x^T \Sigma_X(I+U_X^T\Delta_{\Phi}U_X)\Sigma_Xx}{x^T \Sigma_X^2 x}\nonumber\\
&=1+\frac{x^T \Sigma_XU_X^T\Delta_{\Phi}U_X\Sigma_Xx}{x^T \Sigma_X^2 x}.
\end{align*}
We will be interested in the range of values that the fraction $(x^T M x)/(x^T \Sigma_X^2 x)$ can take over all nonzero $x \in \mathbb{R}^k$. We note that for any vector $x \in \mathbb{R}^k$ we can associate a vector $y:=\Sigma_X x\in\mathbb{R}^k$ and write
\begin{equation*}
\frac{x^T M x}{x^T \Sigma_X^2 x}=1+\frac{y^TU_X^T\Delta_{\Phi}U_Xy}{y^T y}.
\end{equation*}
To bound the range of values that this quantity can take, it suffices to consider all vectors $y \in \real^k$ with unit norm. Thus, we focus on the quantity
\begin{align}
\label{eqn:beforeDPSS}
1+y^TU_X^T\Delta_{\Phi}U_Xy&=1+y^TU_X^T(\Phi^T\Phi - I)U_Xy\nonumber\\
&=y^TU_X^T\Phi^T\Phi U_Xy=\|\Phi U_X y\|_2^2.
\end{align}
Our next step is to apply Lemma~\ref{lem:DPSS} on the subspace $\mathcal{X}=\text{colspan}(U_X)$, using the fact that $\|U_X y\|_2=\|y\|_2=1$.
This tells us that with a probability of at least $1-\delta$,
\begin{equation*}
1-\epsilon\leq\|\Phi U_X y\|_2^2\leq1+\epsilon,
\end{equation*}
holds for all unit norm vectors $y\in\mathbb{R}^k$.
Combining this inequality with~\eqref{eqn:beforeDPSS} we get
\begin{equation}
\label{eqn:offdiagonal}
-\epsilon\leq y^TU_X^T\Delta_{\Phi}U_Xy\leq\epsilon,
\end{equation}
which implies that for any nonzero $x \in \real^k$,
\begin{equation}
1-\epsilon \leq\frac{x^T M x}{x^T \Sigma_X^2 x}\leq 1+\epsilon.
\label{eqn:rayleigh}
\end{equation}

In order to complete the proof we use the following lemma, which is a simplification of Lemma~1 in~\cite{Barlow80computingaccurate}.
\begin{lemma}[Lemma 1,~\cite{Barlow80computingaccurate}]
Let $H$ be a diagonal matrix and suppose $\delta H$ has the property that for all nonzero $x$,
$$
g_l\leq\frac{x^T(H+\delta H)x}{x^T H x}\leq g_u,
$$
where $0<g_l\leq g_u$. Then
$$
g_l\leq\frac{\lambda_i(H+\delta H)}{\lambda_i(H)}\leq g_u
$$
for all $i$, where $\lambda_i(Z)$ denotes the $i$th eigenvalue of the matrix $Z$.
\label{lem:sddrayleigh}
\end{lemma}
Applying Lemma~\ref{lem:sddrayleigh} to \eqref{eqn:rayleigh} with $H=\Sigma_X^2$ (which is diagonal) and $H + \delta H = M$ completes the proof of Statement 1 of Theorem~\ref{thm:mainResult02} and also implies that $\text{rank}(Y) = k$.

\subsubsection{Proof of Statement 2}
\label{sec:svecsproof}

In order to prove Statement 2 of Theorem~\ref{thm:mainResult02}, we require the following important theorem.
\begin{theorem}[Theorem~1, \cite{Mathias1998}]
\label{thm:mathias1998}
Let $H=U\Gamma U^{*}$ and $\widetilde{H}=H+\delta H=\widetilde{U}\widetilde{\Gamma}\widetilde{U}^{*}$ be $p\times p$ positive definite matrices. Assume that $U$ and $\widetilde{U}$ are unitary and that $\Gamma=\text{diag}(\gamma_1,\dots,\gamma_p)$ and $\widetilde{\Gamma}=\text{diag}(\widetilde{\gamma}_1,\dots,\widetilde{\gamma}_p)$ are diagonal. Let $S=U^{*}\widetilde{U}$, and assume
\begin{equation*}
\eta=\|H^{-\frac{1}{2}}\delta H H^{-\frac{1}{2}}\|<1,
\end{equation*}
where $H^{-\frac{1}{2}}=U\Gamma^{-1/2}U^{*}$.
Then for any $j$ and for \textit{any} set $\mathscr{T}$ not containing $j$ we have, %
\begin{equation*}
\left( \sum_{i\in\mathscr{T}} |s_{ij}|^2 \right)^{1/2}\leq
\min\left\{1, ~
\max_{i\in\mathscr{T}} \frac{\gamma_i^{1/2}\widetilde{\gamma}_j^{1/2}}{|\gamma_i-\widetilde{\gamma_j}|}\frac{\eta}{\sqrt{1-\eta}}\right\},
\end{equation*}
and, in particular, for any $i\neq j$,
\begin{equation*}
|s_{ij} |\leq \min\left\{1, ~\frac{\gamma^{1/2} \widetilde{\gamma}_j^{1/2} }{|\gamma_i-\widetilde{\gamma}_j|}\frac{\eta}{\sqrt{1-\eta}}\right\}.
\end{equation*}
\end{theorem}

To prove Statement 2, we continue from the proof of Statement 1. In particular, we suppose~\eqref{eqn:offdiagonal} holds for all unit norm vectors $y \in \real^k$ (recall that this event happens with probability at least $1-\delta$). Our first goal will be to prove that this implies that
\begin{equation}
\label{eq:unorm}
\|U_X^T\Delta_{\Phi}U_X\|_2\leq \epsilon.
\end{equation}
To see why this follows, let us for notational simplicity denote $A=U_X^T\Delta_{\Phi}U_X$.
Note that $A$ is a symmetric matrix.
Then,~\eqref{eqn:offdiagonal} says that $-\epsilon\leq y^T A y\leq\epsilon$ holds for all unit norm vectors $y \in \real^k$. Note that this is equivalent to $-\epsilon\leq \frac{y^T A y}{y^T y}\leq\epsilon$.
This fraction is the well known Rayleigh quotient.
It can be shown that the range of values that the Rayleigh quotient takes is confined between the minimum and the maximum eigenvalues of $A$.
Let us denote the maximum and minimum eigenvalues of $A$ as $\alpha_{\max}$ and $\alpha_{\min}$, respectively.
Since equation~\eqref{eqn:offdiagonal} says that the Rayleigh quotient is in between $-\epsilon$ and $\epsilon$ we can infer that $-\epsilon\leq\alpha_{\min}\leq\frac{y^T A y}{y^T y}\leq\alpha_{\max}\leq\epsilon$.
Thus, $\|A\|_2=\max\{|\alpha_{\min}|, |\alpha_{\max}|\}\leq\epsilon$ and so we have proved that (\ref{eq:unorm}) holds.

To quantify $\|v_j-v_j'\|_2$, we look at a different yet equivalent quantity that may simplify the problem.
Let us again look at the matrix $M$ that we introduced in the proof of Statement 1.
We have seen that there is a close connection between the eigenvalues of $M$ and those of $Y^T Y$.
We now show that in order to prove that the eigenvectors of $Y^T Y$ approximate those of $X^T X$, it suffices to study the eigenvectors of $M$.

Remembering that
$$
M=V_X^T Y^TYV_X=V_X^T V_Y \Sigma_Y^2 V_Y^TV_X,
$$
we can see that the eigenvectors of $M$ are closely related to the right singular vectors of $X$ and $Y$.
Specifically, we have that $V_M=V_X^T V_Y$, and denoting the $j$th eigenvector of $M$ as $\widetilde{v}_j$, it is easy to see that $\widetilde{v}_j=V_X^T v_j'$.
This implies that $\langle \widetilde{v}_j, e_j \rangle=\langle v_j, v_j' \rangle$ for $j=1,\dots,k$, where $e_j$ represents the $j$th canonical basis vector.
Furthermore, we note that $\text{colspan}(V_Y) = \text{rowspan}(Y) = \text{rowspan}(X)$ since every row in $Y$ is a linear combination of the rows in $X$ and since we have argued above that $\text{rank}(Y) = \text{rank}(X)$. From this (and the fact that $v_j' \in \text{colspan}(V_Y)$) it follows that $\|\widetilde{v}_j\|_2=\|V_X^T v_j'\|_2=1$.
Now, using the relation $\langle \widetilde{v}_j, e_j \rangle=\langle v_j, v_j' \rangle$ and the facts that $\|\widetilde{v}_j\|_2=\| e_j \|_2=\|v_j\|_2=\|v_j'\|_2=1$, we see that $\|\widetilde{v}_{j}-e_j\|_2=\|v_j-v_j'\|_2$.
To make sense out of the quantity $\|\widetilde{v}_{j}-e_j\|_2$, let us examine the expression $M=\Sigma_X^2+\Sigma_X U_X^T\Delta_{\Phi}U_X\Sigma_X$.
We can view $M$ as the sum of a diagonal matrix $\Sigma_X^2$ and a perturbation matrix $\Sigma_X U_X^T\Delta_{\Phi}U_X\Sigma_X$.
The eigenvectors of $\Sigma_X^2$ are the canonical basis vectors $e_j$.
Therefore, the quantity $\|\widetilde{v}_{j}-e_j\|_2$ reflects the amount of perturbation in the eigenvectors of $M$.
This is why, to bound $\|v_j-v_j'\|_2$, it suffices to focus on the perturbation analysis of $M$.

We apply Theorem~\ref{thm:mathias1998} as follows:
As we have discussed, we will quantify $\|v_j-v_j'\|_2$ via $\|\widetilde{v}_j-e_j\|_2$.
Let us set the original unperturbed matrix as $H=\Sigma_X^2$ and the perturbation to this matrix as $\delta H=\Sigma_X U_X^T\Delta_{\Phi} U_X\Sigma_X$, such that 
\begin{equation*}
\widetilde{H}=H+\delta H=\Sigma_X(I+U_X^T\Delta_{\Phi} U_X)\Sigma_X=M,
\end{equation*}
and both $H$ and $\widetilde{H}$ are $k\times k$.
Clearly, $H$ is positive definite since it is a diagonal matrix with all positive entries along the diagonal (because $\text{rank}(X) = k$).
To check that $M$ is positive definite, note that $M=\Sigma_X U_X^T\Phi^T\Phi U_X\Sigma_X$ is of the form $M=B^T B$, where $B=\Phi U_X\Sigma_X$ is an $m\times k$ matrix with $m\geq k$.
The fact that $B$ has full column rank will follow because all diagonal entries of $\Sigma_X$ are nonzero (again, because $\text{rank}(X) = k$) and because in the proof of Statement 1 we applied Lemma~\ref{lem:DPSS} on the subspace $\mathcal{X}=\text{colspan}(U_X)$.
Because $B$ has full column rank, $M$ will be positive definite.

We further have that
\begin{equation*}
\eta=\| H^{-\frac{1}{2}}\delta H H^{-\frac{1}{2}} \|_2=\| \Sigma_X^{-1} \delta H \Sigma_X^{-1} \|_2= \|  U_X^T\Delta_{\Phi} U_X   \|_2.
\end{equation*}
Then, applying~\eqref{eq:unorm}, $\eta=\|  U_X^T\Delta_{\Phi} U_X   \|_2\leq\epsilon$.
Let us set $S=I^{T} V_M=V_M$ and denote the $j$th eigenvalue of $M$ as $\widetilde{\lambda}_j$.
Then, straightforward application of Theorem~\ref{thm:mathias1998} yields
\begin{align*}
\left( \sum_{i\neq j} |s_{ij}|^2 \right)^{1/2}&\leq
\min\left\{1,~\max_{i\neq j} \frac{\sigma_i\widetilde{\lambda}^{1/2}_j}{|\sigma^2_i-\widetilde{\lambda}_j|}\frac{\eta}{\sqrt{1-\eta}}\right\},\nonumber\\
&\leq
\min\left\{1,~\max_{i\neq j} \frac{\sigma_i\widetilde{\lambda}^{1/2}_j}{|\sigma^2_i-\widetilde{\lambda}_j|}\frac{\epsilon}{\sqrt{1-\epsilon}}\right\}.
\end{align*}
As we have discussed in Section~\ref{sec:problemsetup} we assume that $s_{jj}=\langle \widetilde{v}_j,e_j\rangle=\langle v_j,v_j'\rangle\geq 0$.
Then,
\begin{align*}
\|v_j-v'_j\|_2&=\|\widetilde{v}_j-e_j\|_2=\sqrt{\|\widetilde{v}_j\|_2^2-2\langle \widetilde{v}_j,e_j\rangle +\|e_j\|_2^2},\nonumber\\
&=\sqrt{2}\sqrt{1-s_{jj}},\nonumber\\
&=\sqrt{2}\sqrt{1-\sqrt{1-\sum_{i\neq j} |s_{ij}|^2}},\nonumber\\
&\leq\sqrt{2}\sqrt{1-\sqrt{1-\min\left\{1,~ \max_{i\neq j} \frac{\sigma^2_i\widetilde{\lambda}_j}{(\sigma^2_i-\widetilde{\lambda}_j)^2}\frac{\epsilon^2}{1-\epsilon}\right \}}  },\nonumber\\
&\leq \sqrt{2}\min\left\{1,~\max_{i\neq j} \frac{\sigma_i\widetilde{\lambda}^{1/2}_j}{|\sigma_i^2-\widetilde{\lambda}_j|}\frac{\epsilon}{\sqrt{1-\epsilon}}\right\},
\end{align*}
where the last inequality is due to the fact that $1-\sqrt{1-x}\leq x$ for $0\leq x\leq 1$.

To write the above only in terms of the unperturbed singular values, $\sigma_j$, we make use of Statement 1 and the fact that $\widetilde{\lambda}_j=(\sigma_j')^2$ for $j=1,\dots,k$ to obtain
\begin{align*}
&\|v_j-v'_j\|_2\leq\nonumber\\
&\min\left\{\sqrt{2},\frac{\epsilon\sqrt{1+\epsilon}}{\sqrt{1-\epsilon}}\max_{i\neq j} \frac{\sqrt{2}\sigma_i\sigma_j}{\displaystyle\min_{c\in[-1,1]}\{|\sigma^2_i-\sigma^2_j(1+ c\epsilon)|\}}\right\}.
\end{align*}

\section{Application to Spectral Analysis of Streaming Graphs}\label{sec:application}

In this section, we apply our data analysis framework to streaming graphs, a model of data collection where edges of a graph are updated dynamically.
We consider a scenario in which edges are inserted and deleted over an observation period, and our goal is to maintain a small data structure that encodes the graph information so that we may analyze the spectrum of the graph quickly at any point during or after the sequence of edge updates.
Spectral graph analysis has a multitude of applications including graph embedding, graph isomorphism testing, data clustering/segmentation (of which there are yet many more applications!), numerical linear algebra, etc. We refer the reader to just a few in \cite{belkin2001laplacian,ng2002spectral,spielman2009spectral,chung1997spectral}.
Determining the spectrum of a graph is at the heart of many modern data analysis and graphical information processing algorithms. 

Let $G = (V,E)$ be a graph with vertex set $V$ and undirected, unweighted edges $E$.  Let $A$ denote the symmetric binary adjacency matrix of $G$, denote by $d_v$ the degree of a vertex $v \in V$, and define the graph Laplacian as
\[
    L_G(u,v) = \begin{cases}
                  d_v & \text{if $u = v$} \\
                  -1    & \text{if $u$ and $v$ are adjacent} \\
                  0     & \text{otherwise}.
			  \end{cases}
\]
There is a compact definition of $L_G$ using the adjacency matrix: $L_G = {\rm diag}(d_v) - A$.

Let $X$ be the incidence matrix of the graph $G$.  This matrix has $N = |E|$ rows and $n = |V|$ columns and to define each entry of $X$, consider an edge $(u,v)$ between vertices $u$ and $v$. Since the graph is undirected, the ordering of the vertices is chosen arbitrarily. Then,
\[
	X_{(u,v),u} = 1 \quad\text{and}\quad X_{(u,v),v} = -1.
\]
It is well-known that the rank of the graph is the rank of the incidence matrix and that this value is $|V| - c$ where $c$ is the number of connected components in $G$.
If the graph $G$ is weighted, we replace the $\pm 1$'s with the appropriate weights in the incidence matrix.

From the definitions of the graph Laplacian and the incidence matrix, it is clear that $L_G = X^T X$. The singular values of $X$ are, therefore, related to the eigenvalues of $L_G$ in a straightforward fashion:
\[
	\sigma_i(X) = \sqrt{\lambda_i(L_G)}.
\]
Furthermore, the {\em right} singular vectors $V$ of $X = U_X \Sigma_X V_X^T$ are the eigenvectors of the Laplacian. Thus, it is sufficient to compute (good) approximations to the singular values and the right singular vectors of $X$ to obtain (good) approximations to the top eigenvalues of $L_G$ and the corresponding eigenvectors. From standard spectral graph theory, we know that the eigenvalues $\lambda_i$ of the Laplacian satisfy $\lambda_1 \geq \cdots \geq \lambda_n = 0$ and, with the assumption that $G$ has $c$ connected components,
$$
\lambda_1 \geq \cdots \geq \lambda_{n-c} > \lambda_{n-c+1} = \cdots = \lambda_n = 0.
$$
In particular, $\text{rank}(X) = \text{rank}(L_G) = n-c$.

Next, we define the streaming graph model. Following~\cite{Ahn:Guha:McGregor1,Ahn:Guha:McGregor2}, we define a dynamic graph stream as a stream of edge updates (both insertions and deletions). This is a faithful model of the evolution of an online social network, for example, in which users connect and disconnect to other users over time \cite{KumarNovakTomkins:2010}.
\begin{definition}[Dynamic graph stream]
	A stream $S = \langle a_1,\ldots, a_T \rangle$ where $a_t = (j_t,k_t,\Delta_t) \in [n] \times [n] \times {\mathbb R}$ defines a weighted graph $G = (V,E)$ where $V=[n]$ and the weight of an edge
$(j,k)$ is given by
\[
	A(j,k) = \sum_{t:(j_t,k_t) = (j,k)~\text{or}~(k,j)} \Delta_t.
\]
We assume that at any update time $t$, the adjacency matrix $A$ is well-formed; that the edge
weight is non-negative; and that the graph has no self-loops.
\end{definition}

In this prototype application, the stream of edge updates $S$ defines the edge-vertex incidence matrix $X$ of the graph $G$.  The matrix $X$ has $N = \binom{n}{2}$ rows and $n$ columns, and for each stream item, we update two entries in $X$ as
\begin{align*}
X_{(j_t,k_t),j_t} & = X_{(j_t,k_t),j_t} + \Delta_t, \\
X_{(j_t,k_t),k_t} & = X_{(j_t,k_t),k_t} - \Delta_t.
\end{align*}
We collect sketches of each column of $X$ and aggregate them into a matrix $Y$. Denoting the $j$th column of $X$ by $x_j$, the $j$th column of the sketched matrix is given by $y_j = \Phi x_j$. We can update the sketch
in a streaming fashion.  Upon receipt of a stream item $(u_t,v_t,\Delta_t)$, we update $y_{u_t}$ and $y_{v_t}$:
\begin{align*}
y_{u_t} &= y_{u_t} + \Delta_t \phi_{(u_t,v_t)}, \\
y_{v_t} &= y_{v_t} - \Delta_t \phi_{(u_t,v_t)},
\end{align*}
where $\phi_j$ denotes the $j$th column of $\Phi$.

Our main result, Theorem~\ref{thm:mainResult02}, tells us that a sketch of the matrix $X$ is sufficient to recover information about its singular value decomposition.
\begin{corollary}
Assume that the undirected, weighted graph $G = (V,E)$ is presented in a streaming fashion so that its incidence matrix $X$ has $n = |V|$ columns, $N = \binom{n}{2}$ rows, and rank $k \le n-1$.
Let $\epsilon\in (0,1)$ denote a distortion factor and $\delta\in(0,1)$ denote a failure probability, and suppose $\Phi$ is an $m \times N$ random matrix that satisfies the distributional JL property with
\begin{equation*}
m\geq \frac{k\log(42/\epsilon)+\log(2/\delta)}{f(\epsilon/\sqrt{2})}.
\end{equation*}
Let $Y = \Phi X$ denote an $m \times n$ sketch of $X$ maintained in the streaming graph model, and let $\widehat{\Sigma}_X = \Sigma_Y$ and $\widehat{V}_X = V_Y$ denote the estimated singular vectors and right singular values of $X$ returned by Algorithm~\ref{algorithm}.
Then with probability at least $1-\delta$, the following statements hold:
\begin{enumerate}
\item (Preservation of eigenvalues) For all $j = 1,\dots,k$,
\[
	1 - \epsilon \leq \frac{\lambda_j'}{\lambda_j} \leq 1 + \epsilon
\]	
where $\lambda_j$ denote the true eigenvectors of the graph Laplacian $L_G$ and $\lambda_j'$ denote the estimated eigenvalues obtained by squaring the diagonal entries of $\widehat{\Sigma}_X$.
\item (Preservation of eigenvectors) For all $j = 1,\dots,k$,
\begin{align*}
&\|v_j-v_j'\|_2\leq\nonumber\\
&\min\left\{\sqrt{2}, ~ \frac{\epsilon\sqrt{1+\epsilon}}{\sqrt{1-\epsilon}}\max_{i\neq j}\frac{\sqrt{2}\lambda_i^{1/2}\lambda_j^{1/2}}{\displaystyle\min_{c\in[-1,1]}\{|\lambda_i-\lambda_j(1+c\epsilon) |\}  } \right\},
\end{align*}
where $v_j$ are the eigenvectors of the graph Laplacian $L_G$ and $v_j'$ denote the estimated eigenvectors obtained from the columns of $\widehat{V}_X$. 
\end{enumerate}
\end{corollary}
Because the adjacency matrix $A$ of $G$ has at most $|V|^2$ non-zero entries, this result is useful only when the rank $k$ of $L_G$ is significantly smaller than $n = |V|$, the number of vertices; or, equivalently, when $G$ has many connected components.  In this case, the size of the sketch is smaller than that of the adjacency matrix. In summary, for highly disconnected graphs presented in a streaming fashion, we can recover the approximate eigenvalues and eigenvectors of the Laplacian.  The sparsity of the matrix $\Phi$ and the speed with which we can update the sketch matrix $Y$ under a stream of updates are functions of the quality of the JL transform. The structural evolution of online social networks \cite{KumarNovakTomkins:2010} suggests that it is reasonable to assume that the underlying graph has a significant fraction of vertices in small, disconnected components so that the graph is essentially a low-rank graph.

\section{Conclusion}\label{sec:conclusion}

We present a data collection and analysis scheme that permits the distributed collection of data $X$ by resource constrained sensors in a network and the central computation of the spectral decomposition of $X^T X$ or the right singular vectors of the data itself.  
The algorithm returns not an approximation to the original data, but a good approximation to the singular values $\sigma_j$ and the right singular vectors $v_j$ of the data. 
This data collection and analysis framework makes a small number of linear, non-adaptive measurements of the data. 
The number of measurements each sensor makes is comparable to the rank of the data and, if the data are full rank, the number of measurements at each sensor is comparable to the total number of sensors.
This efficient data collection is especially important for sensors that are severely resource constrained and cannot store or transmit a large amount of data to a central device; we believe that one possible application of such an algorithm would be in operational modal analysis of structures (buildings, bridges, etc.)~\cite{peeters:659}.

\section{Acknowledgments}

This work was partially supported by NSF grant CIF-0910765, AFOSR grant FA9550-09-1-0465, and NSF CAREER grant CCF-1149225. The authors thank Jerome Lynch and Sean O'Connor, University of Michigan, for helpful discussions about modal analysis in structural health monitoring.

\bibliographystyle{abbrv}
\bibliography{refs,annarefs}
\end{document}